# Development of a 3-D energy-momentum analyzer for meV-scale energy electrons

S. Karkare,[1, a)] J. Feng,[2] J. Maxson,[3] and H. A. Padmore[4]
[1)] *Department of Physics, Arizona State University, Tempe, AZ-85282*
[2)] *Lawrence Berkeley Lab, Berkeley, CA, 94520*
[3)] *Cornell Laboratory for Accelerator-Based Sciences and Education, Cornell University, Ithaca, New York 14853, USA*
[4)] *Advanced Light Source Division, LBNL, Berkeley, California 94720, USA*

(Dated: 13 February 2019)

In this article, we report on the development of a time-of-flight based electron energy analyzer capable of measuring the 3-D momentum and energy distributions of very low energy (meV-scale) photoemitted electrons. This analyzer is capable for measuring energy and 3-D momentum distributions of electrons with energies down to 1 meV with a sub-meV energy resolution. This analyzer is an ideal tool for studying photoemission processes very close to the photoemission threshold and also for studying the physics of photoemission based electron sources.

## I. INTRODUCTION

Over the past several decades, the study of photoelectron energy and momentum distributions has revealed to us a great deal regarding the chemistry and physics of materials[1]. Today, commercially available angle resolved electron energy analyzers are capable of measuring the energy and angular distributions (or equivalently, the 3-D momentum distributions) with sub-meV energy resolutions and are routinely used for X-ray Photoelectron Spectroscopy (XPS) or Angle Resolved Photoelectron Spectroscopy (ARPES) to deduce the chemical composition and electronic structure of materials. Such analyzers either use concentric hemispheres as energy filters and image the emission angles of the electrons on a screen[2] or utilize a series of electron lenses with a delay-line-detector to measure the time-of-flight[3] to deduce the electron energies and momenta. Typically, these analyzers are designed to study the energy and momentum distributions of electrons with (kinetic) energies ranging from $\sim$ 1 eV to a few keV[4].

Usually, in XPS or ARPES applications the photon energies used to emit electrons are much larger than the work-function and hence, the energies of the photoelectrons emitted are of the same order of magnitude as the photon energies ($\sim$ few 10s of eV to few keV)[1]. More recently, with the advent of laser-based-ARPES, energy-momentum distributions of photoelectrons emitted with photon energies only a few eV larger than the work function have been investigated[5]. The photoelectrons emitted and detected in such laser-based-ARPES techniques have energies ranging from several 100 meV to a few eV. Thus, commercially available analyzers are sufficient to study the range of electron energies of interest to all well-developed XPS and ARPES applications.

Photoemission is also widely used as a source of electrons for ultra-fast electron diffraction and microscopy[6] and linear particle accelerator applications like Energy Recovery Linacs[7] and Free Electron Lasers[8]. The brightness of electron beams is a key figure of merit that determines the performance of such applications. With significant advancements in beam shaping and space-charge management techniques in the past several decades, today the beam brightness is often limited by the mean transverse energy (or equivalently the rms transverse momentum) of the electrons photoemitted from the cathode[9]. Here, the transverse direction is the direction perpendicular to the direction of electron beam propagation (or the direction parallel to the cathode surface). A smaller mean transverse energy (MTE) implies brighter electron beams. Hence, in the past decade a lot of effort has gone into studying the photoemission process and minimizing the MTE obtained from photocathodes. Often, reducing the MTE is a trade off between other crucial performance metrics of photocathodes like the quantum efficiency (QE) and response time[10,11].

The smallest MTE's are obtained when the photon energy is very close to the photoemission threshold and the energies of the emitted electron are near zero (up to a couple of 100 meV). Measuring and studying the energy and momentum distributions of such low energy electrons is crucial to understanding the photoemission process at photon energies very close to the photoemission threshold and develop cathode materials that can minimize the MTE without significantly impacting other crucial photocathode metrics. Such low energy electrons are very slow and have long (up to 15 nm) de Broglie wavelengths. Hence the physics of photoemission of these electrons can be very different from the higher energy electrons typically used for photoelectron spectroscopy. Therefore, developing an analyzer capable of measuring energy and 3-D momentum distributions of very low energy (sub-100 meV) electrons is essential. Commercially available energy-momentum analyzers used for photoelectron spectroscopy are not designed to measure such low energy electrons. Furthermore, such analyzers only accept electron emitted in a fairly narrow cone about the surface normal and cannot measure the electrons emitted at large angles. It is crucial to measure these electrons due to their significant contribution to the MTE.

One challenge in measuring such low energy electrons is the sensitivity of their trajectories to stray magnetic and

---

a)Electronic mail: karkare@asu.edu

electric fields. In the past, several techniques that minimize the sensitivity to stray fields have been developed to measure the transverse and longitudinal energy distributions of such low energy electrons. So far these techniques have been successfully implemented to either measure the 2-D transverse momentum distributions[12,13], 2-D longitudinal and transverse energy distributions[14,15] or the 1-D longitudinal energy distributions[16].

In this paper, we report of the development of a technique to measure the complete 3-D energy and momentum distributions of very low energy electrons using the time-of-flight approach. Using this approach we have developed an electron energy analyzer that can measure these distributions with a sub-meV expected resolution. In section 2 we discuss the principle of the time-of-flight based measurement along with the design of the analyzer and the various contributions to its resolution and possible sources of error. In section 3 we show the measurements of the surface state on the (111) surface of silver as an example and discuss the calibration and operational details of the analyzer. A techniques similar to the one presented below was developed by Kirchmann et al. However, their implementation did not allow proper measurement of very low energy electrons and was capable of detecting electrons emitted only in a 22 degree cone about the normal[17].

## II. DESIGN OF THE TOF-BASED ANALYZER

### A. Basic design and principle of measurement

Figure 1a shows the cross-section of the analyzer. The analyzer has a parallel plate geometry, with one plate being the sample (photoemission surface) along with an electrostatic shield electrode and the other being the time-of-flight detector. The electrostatic shield is a cylindrical disk with a central hole into which the sample is inserted. Note that the sample is slightly recessed from the electrostatic shield due to practical considerations as seen in figure 1a. The electrostatic shield and the sample are isolated electrically. The electrostatic shield is uniformly coated with a graphite film to obtain a uniform work function, The voltage applied to the electrostatic shield roughly is equal to the voltage applied to the sample plus the work function difference between the sample and the shield. This causes the electric field in the region between the sample and the detector to be uniform. The detailed procedure for obtaining the uniform electric field is discussed in section 3. A small accelerating voltage of a few volts ($\sim$ 0-10 V) is applied between the sample and the detector. The whole setup is placed in an ultra high vacuum chamber with walls made of mu-metal to minimize the magnetic fields within the chamber. Combined with non magnetic construction the fields at the sample are only a few milli-Gauss. The sample is also thermally connected to a liquid Helium (LHe) cryostat to enable cooling to cryogenic temperatures. Figure 1b shows the 3-D model of the analyzer.

A fs to ps scale laser pulse is focused into a spot of size less than 100 $\mu$m FWHM on the sample center. The intensity of the laser pulses is kept small enough that no more than one electron is emitted per pulse. Due to the uniform electric field, the emitted electrons are accelerated only in the longitudinal direction towards the detector and do not experience any significant transverse ($x$ or $y$, along the direction parallel to the sample surface) acceleration (more details discussed in section 2d). The detector detects the $x$ and $y$ position of the electrons and the time required by the electrons to travel to the detector ($t$). The laser spot size on the sample is much smaller than the spot formed by the electrons on the detector and hence we can assume that the electrons are emitted from a single point on the sample. From these measurements, momenta of the emitted electrons ($\hbar k_x$, $\hbar k_y$ and $\hbar k_z$) can be calculated as

$$\hbar k_x = m_e \frac{x}{t} \tag{1}$$

$$\hbar k_y = m_e \frac{y}{t} \tag{2}$$

$$\hbar k_z = m_e \left[ \frac{d}{t} + \frac{(V - V_{\text{off}})\, et}{2 m_e d} \right] \tag{3}$$

and the total energy can be calculated as

$$E = E_\perp + E_\parallel \tag{4}$$

where $m_e$ is the electron rest mass, $e$ is the absolute value of the electron charge, $d$ is the effective distance between the sample and the detector, $V$ is the voltage applied to the sample (the detector is grounded) and $V_{\text{off}}$ is the work function difference between the sample and the detector (work function of the sample minus the work function of the detector). $E_\perp = \frac{\hbar^2}{2 m_e}\left(k_x^2 + k_y^2\right) = \frac{\hbar^2 k_r^2}{2 m_e}$ is the transverse energy and $E_\parallel = \frac{\hbar^2 k_z^2}{2 m_e}$ is the longitudinal energy. Here $r = \sqrt{(x^2 + y^2)}$ is the radial coordinate at which the electron hits the detector.

The detector records the $x$, $y$ and $t$ over a period of time for many electrons in order to generate a distribution $N(x, y, t)\, dx dy dt$. This distribution can be converted into a distribution over the total energy $E$ and the two transverse momenta $k_x$ and $k_y$ as $N(k_x, k_y, E)\, J_{k_x, k_y, E} dk_x dk_y dE$, where $J_{k_x, k_y, E}$ is the determinant of the appropriate Jacobian matrix and is given by

$$J_{k_x, k_y, E} = \frac{1}{\begin{vmatrix} \frac{\delta E}{\delta t} & \frac{\delta E}{\delta x} & \frac{\delta E}{\delta y} \\ \frac{\delta k_x}{\delta t} & \frac{\delta k_x}{\delta x} & \frac{\delta k_x}{\delta y} \\ \frac{\delta k_y}{\delta t} & \frac{\delta k_y}{\delta x} & \frac{\delta k_y}{\delta y} \end{vmatrix}} \tag{5}$$

.

From equations 1 and 2 we can see that the transverse momenta depend only on the measured values of $x$, $y$ and $t$ and do not require the knowledge of the distance $d$ and the work function difference $V_{\text{off}}$. However, calculating the longitudinal momentum $\hbar k_z$ and the total energy $E$ requires the accurate knowledge of $d$ and $V_{\text{off}}$. Obtaining these values accurately requires the use of some known feature in the energy or longitudinal momentum distribution like the energy of a surface state or the Fermi level. An example method of estimating the values of $d$ and $V_{\text{off}}$ using the surface state of Ag(111) has been shown in section 3.

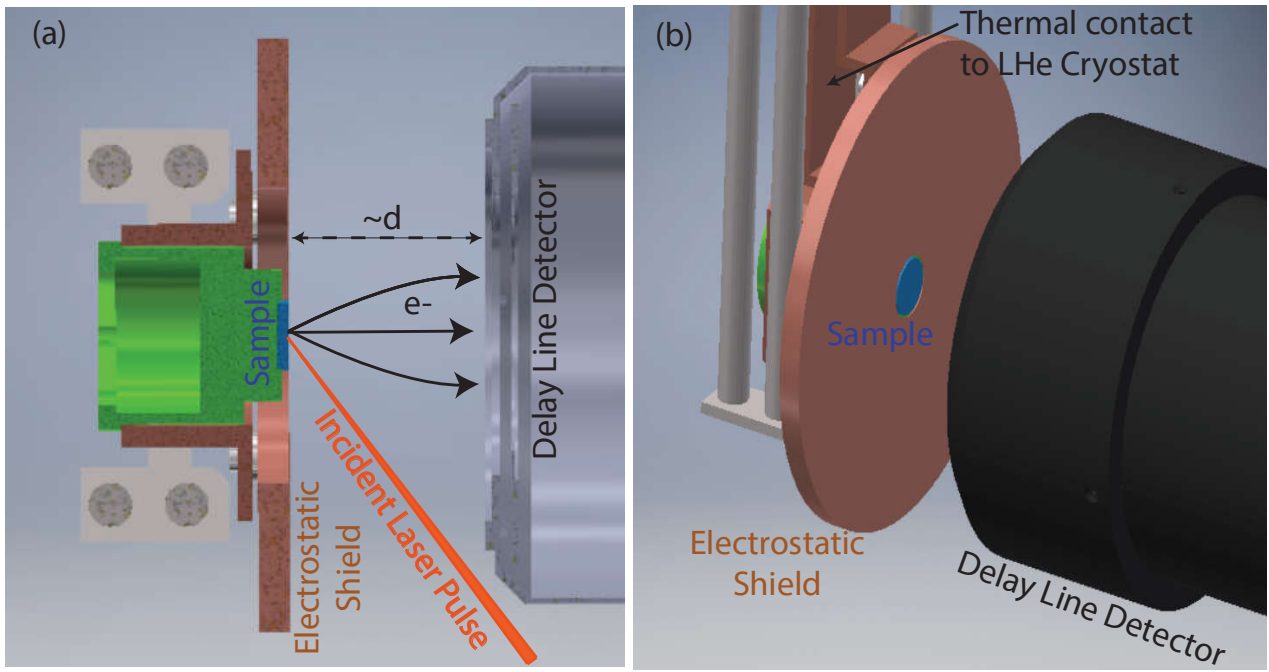

FIG. 1. (a) Cross-section and (b) 3-D model of the analyzer showing the parallel plate configuration formed by the sample, electrostatic shield and the detector.

### B. Detector and the time-of-flight measurement

A time resolved imaging MCP detector based on delay line detection (DLD4444 manufactured by Surface Concepts[18]) is used to detect the electrons and measure $x$, $y$ and $t$. The detector has a spatial resolution better than 40 $\mu$m and an intrinsic time resolution better than 30 ps. In practice, due to binning, the spatial resolution used for the measurements presented in this paper is 0.11 mm in the $x$ direction and 0.12 mm in the $y$ direction. The spatial resolution is also limited by the laser spot size on the sample to 100$\mu$m FWHM. The temporal resolution is limited by the timing jitter between the laser and the trigger to the delay line detector to 170 ps FWHM.

The detector has an active area of 40 mm and comprises of a fine grid followed by a MCP. Behind the MCP two meandering delay lines (one for the $x$ direction and the other for the $y$ direction) detect the amplified electron pulse to give the $x$ and $y$ positions and the time $t$ of the electron hitting the MCP. The laser pulse reflected off the sample surface is also detected by the delay-line-detector and is used to set the time of emission as $t = 0$. The temporal width of the reflected laser pulse gives the effective temporal resolution of the detector after including the effect of the jitter. For an isotropic distribution of the emitted electrons, $x = 0$ and $y = 0$ can be set as the location of the centroid of the $x$ and $y$ positions of all the electrons detected.

The detector is capable of reliably detecting at most one count per laser pulse. The laser power incident on the sample must be reduced to ensure this condition.

### C. Energy resolution

The total energy resolution $\delta E$ can be expressed as the sum of the longitudinal and transverse energy resolutions as $\delta E = \delta E_\parallel + \delta E_\perp$.

From equations 1, 2 and 3 we can calculate

$$\delta E_\parallel = [\kappa_d \delta d + \kappa_t \delta t + \kappa_{V_e} \delta V_e] \tag{6}$$

and

$$\delta E_\perp = \kappa_r \delta r + \kappa_{t1} \delta t \tag{7}$$

where $\kappa_d = \sqrt{2E_\parallel m_e d}\left(\frac{1}{t} - \frac{V_e e t}{2 m_e d^2}\right)$, $\kappa_t = \sqrt{2E_\parallel m_e d}\left(-\frac{d}{t^2} + \frac{V_e e}{2 m_e d}\right)$, $\kappa_{V_e} = \sqrt{2E_\parallel m_e d}\left(\frac{e t}{2 m_e d}\right)$, $\kappa_r = \frac{m_e r}{t^2}$, $\kappa_{t1} = \frac{m_e r^2}{t^3}$ and $V_e = V - V_{\text{off}}$. The time $t$ can be given as a function of the longitudinal energy $E_\parallel$ to obtain the terms $\kappa_d \delta d$, $\kappa_t \delta t$ and $\kappa_{V_e} \delta V_e$ as functions of $E_\parallel$, $d$ and $V_e$. The value of $d$ is constrained to approximately 40 mm due to the geometry of the setup. While the value of temporal uncertainty $\delta t$ can be measured easily as the temporal resolution of the detector (see previous section), the possible values of the uncertainties in $d$ ($\delta d$) and $V_e$ ($\delta V_e$) are more difficult to determine. Here we make educated guesses to the values of these uncertainties.

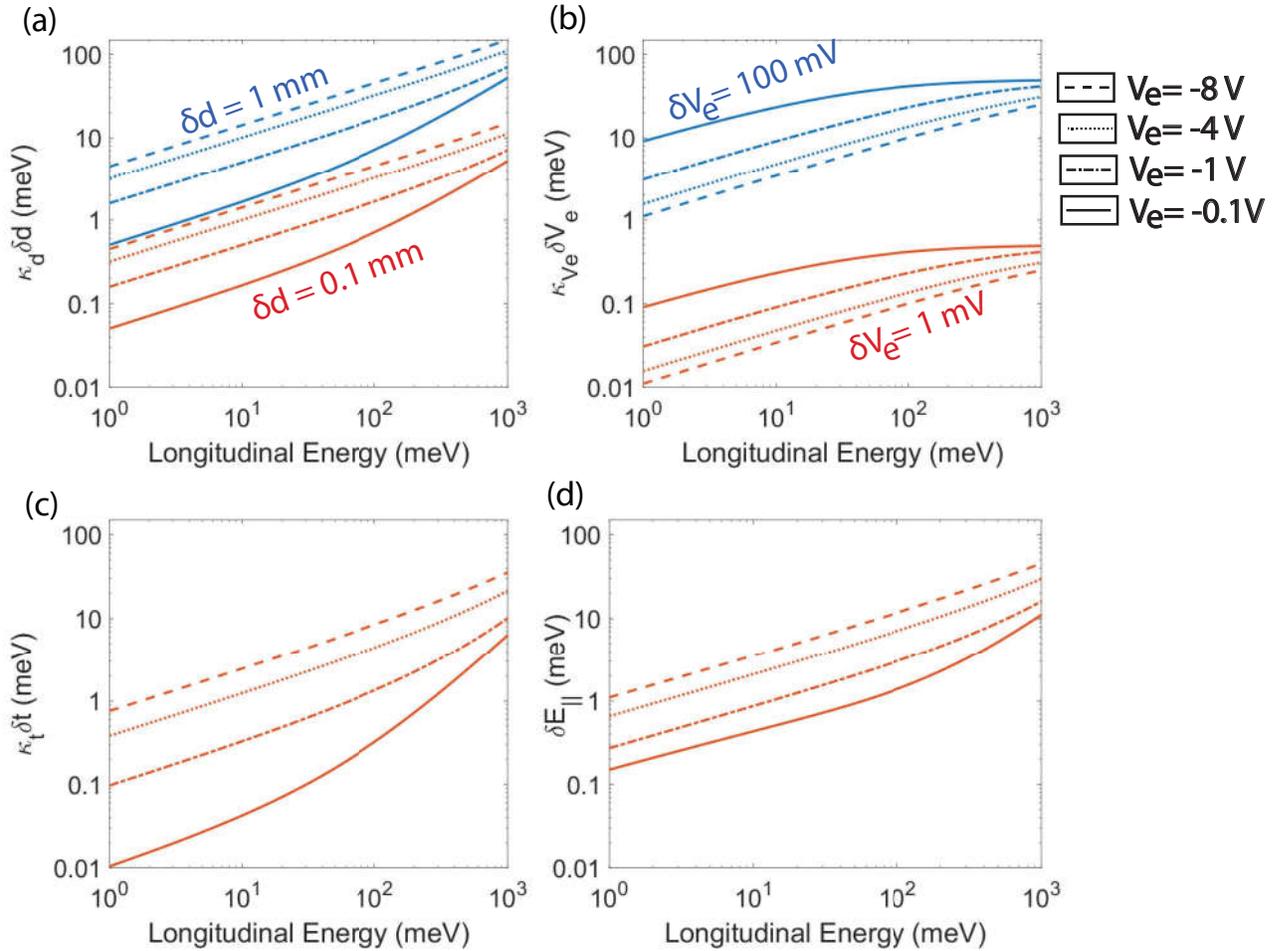

FIG. 2. (a) Contribution of the uncertainty in $d$ to the energy resolution ($\kappa_d \delta d$) for several values of $V_e$. The blue curves correspond to a very large value of $\delta d = 1$ mm, while the red curves correspond to a more realistic values of $\delta d = 0.1$ mm (b) Contribution of the uncertainty in $V$ to the energy resolution ($\kappa_{V_e} \delta V_e$) for several values of $V_e$. The blue curves correspond to a very large value of $\delta V_e = 100$ mV, while the red curves correspond to a more realistic values of $\delta V_e = 1$ mV (c) Contribution of the uncertainty in $t$ to the energy resolution ($\kappa_t \delta t$) for several values of $V_e$ for $\delta t = 200$ ps (d) The total uncertainty in longitudinal energy ($\kappa_d \delta d + \kappa_{V_e} \delta V_e + \kappa_t \delta t$) as a function of the longitudinal energy. Here we assume $\delta d = 0.1$ mm, $\delta V_e = 1$ mV and $\delta t = 200$ ps.

The variation in $d$ arises from a combination of the angle offset between the sample and the detector, vibrations in the system, roughness of the sample and the shield electrode and the surface non-uniformity of the MCP in the detector. Figure 2a shows the contribution of the uncertainty in $d$ to the energy resolution ($\kappa_d \delta d$) for several values of $V_e$. The contribution of the uncertainty in $d$ reduces as the voltage between the sample and the detector $V_e$ is reduced. When $\delta d$ is assumed to be equal to 100 $\mu$m the value of $\kappa_d \delta d$ is less than 1 meV even for longitudinal energies as large as 100 meV at $V_e = -1$ V. Even for very large values of uncertainty in $d$ (as large as 1 mm), the contribution to the energy resolution can be made small by reducing the effective voltage between the sample and the detector.

The variation in $V_e$ can arise from the noise in the voltages supplied to bias sample and the shield electrode or from the work function variations on the sample, shield electrode and detector surfaces. The noise in the bias voltages has been measured to be less than 1 mV. Although care has been taken to maintain the uniformity of the work functions on the surfaces of the shield and the detectors (by coating them with a graphite film), one might still expect work function variations over mm scales due to adsorption of residual gas atoms and impurities limiting the value of $\delta V_e$ to higher than 1 mV. As seen in figure 2b, even for very large variations in the voltage ($\delta V_e = 100$ meV) due to possible work function variations we can get a sub-10 meV energy resolution contribution of voltage uncertainty at 100 meV longitudinal energy at $V_e = -1$ V. For values of $\delta V_e$ limited to 1 meV only by noise in the applied voltages, the contribution of voltage uncertainty is less than 1 meV as shown in figure 2b. The voltage uncertainty contribution to the energy resolution reduces with the effective voltage between the detector and the sample.

The effect of the temporal resolution on the longitudinal energy resolution is shown in figure 2c. We see that for a temporal resolution as large as 200 ps (slightly larger than the measured resolution of 170 ps), at $V_e = -1$ V, the contribution to the energy resolution is about 1 meV.

As seen from figure 2, the uncertainty contributions of $d$ and $t$ to the energy resolution reduce with $V_e$. However, the uncertainty contribution of $V_e$ to energy resolution increases with $V_e$. Hence, there is an optimum voltage to get the maximum energy resolution. The longitudinal energy resolution (assuming $\delta d = 0.1$ mm, $\delta V_e = 1$ mV and $\delta t = 170$ ps) as a function of the longitudinal energy is shown in figure 2d for various values of $V_e$. For the values of the uncertainties assumed here, the longitudinal energy resolution is 3 meV for a longitudinal energy of 100 meV at $V_e = -1$ V and lower at lower energies and voltages. Although this is a realistic uncertainty estimate for the current setup, it can be significantly improved by appropriate mechanical design to minimize vibrations, align the sample, shield and detector; by improving the electrical design to minimize the voltage fluctuations; and by improving the temporal jitter between the laser and the detector. The ultimate resolution will be determined by the temporal resolution of the detector and is below 0.1 meV even for 100 meV longitudinal energy electrons.

If the maximum longitudinal energy is much smaller than $-eV_e$, for the purpose of calculating the resolution we can assume that all electrons start with near zero energies and $t \approx \sqrt{\frac{2m_e}{-V_e e}} d$ for all electrons. Using this value of $t$ we get

$$\kappa_d = \sqrt{2E_\parallel m_e} d \frac{\sqrt{-2V_e e}}{\sqrt{m_e} d} \quad (8)$$

$$\kappa_{V_e} = \sqrt{2E_\parallel m_e} d \frac{\sqrt{e}}{\sqrt{-2V_e m_e}} \quad (9)$$

$$\kappa_t = \sqrt{2E_\parallel m_e} d \frac{V_e e}{2 m_e d} \quad (10)$$

$$\kappa_r = \frac{\sqrt{-E_\perp V_e e}}{d} \quad (11)$$

$$\kappa_{t1} = \frac{2 E_\perp \sqrt{-V_e e}}{\sqrt{2m_e} d} \quad (12)$$

Under the above assumption we see that the longitudinal energy resolution is proportional to the square root of the longitudinal energy. This assumption also allows us to obtain a straight forward, quantitative estimate of the transverse energy and momentum resolution.

Figure 3 gives the transverse energy and transverse momentum resolution. The uncertainty in the radial coordinate is the sum of the effective detector pixel size after binning ($\sim 120\mu$m) and the laser spot size on the sample ($\sim 100\mu$m). The temporal resolution for the calculations in figure 3 is assumed to be 200 ps. We can see that the transverse energy resolution is $\sim 2$ meV for 100 meV transverse energy at $V_e = -1$ V. This translates to a transverse momentum resolution of $\sim 1.5 \times 10^{-3}$ 1/$A^\circ$ at $V_e = -1$ V as seen in figure 3b. This is comparable to the transverse momentum resolution obtained in most ARPES systems and a tiny fraction of a typical Brillouin zone width. Nearly a factor of 3 better resolution can easily be achieved by reducing $\delta r$. This can be done by reducing the laser spot size and by reducing the binning to increase the detector's position resolution.

### D. Effect of stray electric and magnetic fields

It is critical to establish a constant electric field along with a near zero magnetic field in the space between the sample/electrostatic shield and the detector. The mu-metal vacuum chamber acts as a magnetic shield reducing the magnetic field below 10 mG. The sample/electrostatic shield and the detector form a parallel plate capacitor like geometry. The outer diameter of the detector is roughly 2 times larger than the detector sample distance, ensuring sufficiently constant electric field in the region in which electrons travel towards the detector. In this section we obtain a quantitative estimate of the errors produced in the measurement due to the non-uniform electric fields and the stray magnetic fields.

The non-uniformity in the electric field is due to the fringe effects from the finite size of the electrostatic shield and the detector and inaccurate determination of the work function difference between the sample and the electrostatic shield. For assessing the contribution of each of these, the electric fields between the sample and the detector were calculated for various values of the effective voltage between the sample and the detector ($V_e$) using the 2-D electrostatic solver POISSON-Superfish[19]. Figure 4 shows the cross-section of the cylindrically symmetric geometry used for the calculation and the equipotential lines calculated for one example voltage of $V_e = -4$ V.

The sample and the electrostatic shield in general have different work functions producing a significant non-uniformity in the electric fields. If the work function difference is known, an equivalent offset can be added to the voltages applied to the sample and the electrostatic shield. Determining the work function difference accurately is non-trivial and a technique to do so to an accuracy of $\sim 5$ mV has been presented in the next section. To account for this inaccuracy in the determination of the work function difference, a difference of 5 mV between the effective voltages of the sample and the electrostatic shield has been assumed in the calculations of the electric fields.

Trajectories of electrons launched with various longitudinal and transverse energies from the center of the sample were traced in the calculated electric fields to obtain the final transverse position and the time-of-flight to the detector. The longitudinal and the transverse energies were then calculated from the final transverse position and the time of flight using equations 1, 2 and 3 and compared to the initial energies that the electrons were launched with.

Figure 5 shows the difference between the longitudinal energy calculated from the simulated time-of-flight and the initial launch longitudinal energy ($\Delta E_\parallel$) as a function of the initial longitudinal energy for various values of $V_e$. Figure 5a corresponds to electrons tracked in the electric field generated by assuming a 5 mV difference between the effective voltages of the sample and the electrostatic shield, while figure 5b corresponds to electrons tracked in the electric field generated by assuming the effective voltage difference to be 20 mV.

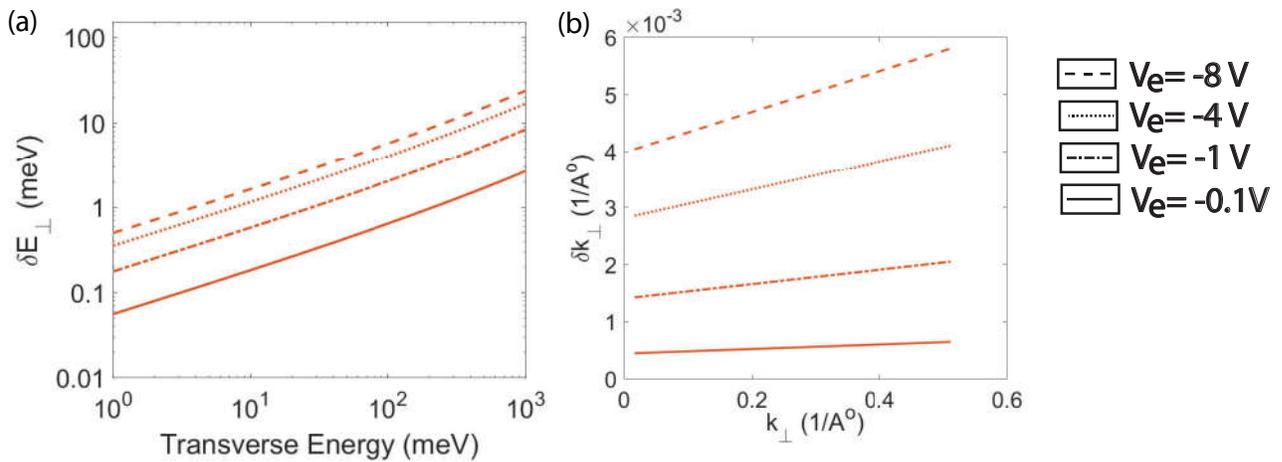

FIG. 3. (a) Uncertainty in transverse energy for various values of $V_e$ (b) Uncertainty in transverse momentum for various values of $V_e$

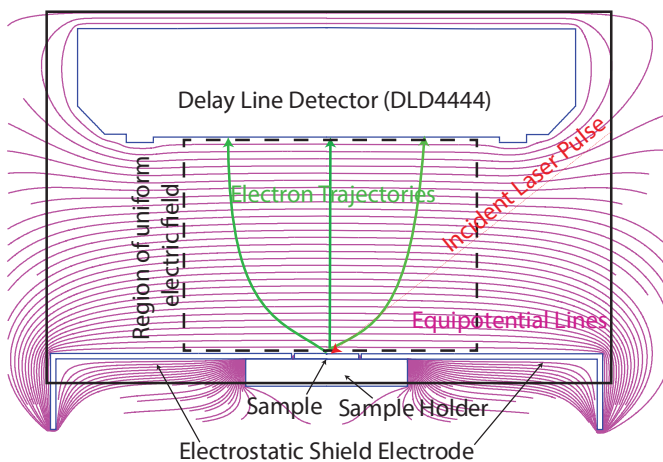

FIG. 4. Cross-section of the energy analyzer showing the equipotential lines between the detector and the sample. Here $V_e = 4$ V and an offset of 5 mV between the sample and the shield voltages.

For higher values of $V_e$ the effects of the fringe distortions due to finite size of the detector and the shield electrode dominate causing $\Delta E_\parallel$ to become larger than 10 meV even at small ($\sim 10$ meV) longitudinal energies as seen in figure 5. For smaller values of $V_e < -2$ V, if the effective voltage difference between the sample and the shield is 5 mV or less, the effects of the fringe distortions are minimal and $\Delta E_\parallel$ is limited to sub-meV values for longitudinal energies as small as 10 meV as seen in figure 5a. However, for larger values of the effective voltage difference (20 mV as seen in figure 5b), $\Delta E_\parallel$ is several meV for longitudinal energy of 10 meV even at small values of $V_e$. It was noted that the values of $\Delta E_\parallel$ did not depend strongly on the transverse energy of the electrons.

Similar calculations were performed for the inaccuracy in the transverse energy and the electric field non-uniformities were found to have an negligible effect on the transverse energy inaccuracy at all values of $V_e$, even for 20 mV difference in the effective voltages of the sample and the shield electrode. It was found that adding a magnetic field of 10 mG in all three directions did not affect the trajectories significantly and produced errors of less than 1% in the calculated energies.

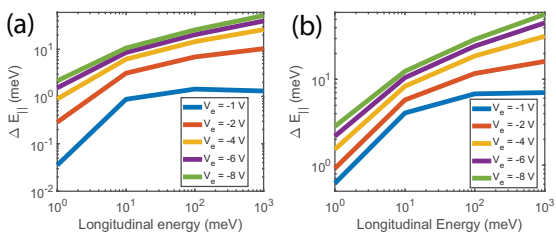

FIG. 5. Expected error in the measured longitudinal energy due to non-uniform electric fields from fringe effects for (a) voltage offset of 5 mV between the sample and the shield and (b) voltage offset of 20 mV between the sample and the shield. The error in longitudinal energy was found to be nearly independent of the transverse energy. The error in transverse energy due to fringe effects was negligible.

The above results suggests that the operation at low values of $V_e$ results in a higher accuracy in the longitudinal energy and reducing the effective voltage difference between the sample and the electrostatic shield is critical. We also see that the field variations do not significantly affect the transverse energy measurement making this a very robust technique to measure the mean transverse energy of photoemitted electrons.

## III. MEASUREMENT OF THE Ag(111) SURFACE STATE

In this section, using the Ag(111) surface as an example we demonstrate the procedure developed to obtain the uniform electric field in the region between the detector and the sample and also the procedure to obtain the exact distance $d$ and voltage offset $V_{\text{off}}$ between the sample and the detector.

A commercially bought Ag(111) single crystal was exposed to several cycles of ion bombardment with 1 keV $Ar^+$ ions and annealing to 450°C. This produces an atomically ordered surface with a Shockley surface state. The cleanliness and atomic ordering of the surface was confirmed using Low Energy Electron Diffraction and Auger Electron Spectroscopy. The work function of this surface is known to be 4.45 eV[11]. This surface was then used to perform the calibrations presented in this section.

Electrons were photoemitted using a frequency tripled Ti-sapphire laser with a tunable photon energy and a pulse length of ∼150 fs. The photon energy could be tuned from 4.2 to 4.9 eV. The laser had a repetition rate of 76 MHz. However, this was reduced by a factor of 20 to 3.8 MHz using an acousto-optic Bragg cell pulse picker to make the repetation rate less that 7 MHz (the maximum trigger rate for the DLD4444 detector). This laser was also used to generate an electrical pules to trigger the DLD4444 detector. During measurement the laser power was adjusted to keep the count rate of electron hitting the detector to less than 50,000 counts per second to avoid damage to the MCP in the detector and ensure that not more than one electron is emitted per laser pulse.

### A. Generating uniform electric field

As shown in the previous section, ensuring that the electric field is uniform between the sample and the detector is crucial to obtain the longitudinal energies and, to a certain extent, essential to calculate the transverse energy/momentum of the emitted electrons from the $x$, $y$ and $t$ measurements. This has been achieved through the use of an electrostatic shield electrode. Ideally, the sample surface and the electrostatic shield surface facing the detector should lie in the same plane. However, due to manufacturing imperfections, it was found that the plane of the sample surface was slightly recessed from the plane of the electrostatic shield surface.

In order to ensure an uniform electric field, the voltages applied to the sample and the electrostatic shield electrode need to be offset. This offset is due to the work function difference between the sample and the electrostatic shield and also due to the fact that the plane of the sample is slightly recessed from the plane of the electrostatic shield. Figure 6a shows a schematic of the geometry with the slightly recessed sample. Let the distance between the plane of the sample surface and the plane of the electrostatic shield surface be $d_1$. The electric field between the sample and the detector will be uniform when the following condition is satisfied:

$$\frac{d - d_1}{d} (V_{\text{sam}} - W_{\text{sam}}) = (V_{\text{sh}} - W_{\text{sh}}). \qquad (13)$$

Here $V_{\text{sam}}$ and $V_{\text{sh}}$ are the voltages applied to the sample and the shield respectively, and $W_{\text{sam}}$ and $W_{\text{sh}}$ are the work functions of the sample and the shield respectively. This relationship can be re-written as:

$$V_{\text{sh}} = K_1 + K_2 V_{\text{sam}} \qquad (14)$$

where $K_1 = -K_2 W_{\text{sam}} + W_{\text{sh}}$ and $K_2 = \frac{d-d_1}{d}$. Since $K_2 \approx 1$, $K_1 = W_{\text{sh}} - W_{\text{sam}}$.

When the condition in equation 13 (and hence also in equation 14) is satisfied, the spot formed by all the electrons hitting the detector does not change position for small changes in the position of the sample+shield assembly in the direction parallel to the plane of the sample surface, as shown in figure 6b. Note that the position of the laser spot is held fixed w.r.t the detector as the sample+shield assembly is moved and electrons are emitted from an off-center location on the sample.

When the condition in equation 13 is not satisfied the electrostatic shield essentially acts like an electron lens. In such a scenario, the electrons emitted from an off-center location on the sample get deflected by the lensing effect of the shield. Thus when the position of the sample+shield assembly changes w.r.t the detector, the electron spot on the detector changes position as shown in figures 6c and 6d.

For a given sample voltage, the shield voltage can be adjusted until the centroid of the electron spot on the detector stops moving with small changes in the position of the sample+shield assembly. This is the shield voltage voltage for which the condition in equation 13 is satisfied. The values of $K_1$ and $K_2$ can be obtained by fitting a line through a plot of various values of $V_{\text{sam}}$ and the corresponding values of $V_{\text{sh}}$. For the case of the Ag(111) surface the values of $K_1 = -375$ mV and $K_2 = 0.99$ were obtained.

### B. Calibrations to calculate the longitudinal energy distributions

Once an uniform electric field is established between the sample and the detector, the transverse momentum and energy can be obtained directly from the $x$, $y$ and $t$ measurements as seen from equations 1 and 2. However, obtaining the longitudinal energy/momentum requires the knowledge of the distance between the sample and the detector $d$ and the voltage offset $V_{\text{off}}$ as evident from equation 3.

From the design, the value of $d$ is expected to be around 40 mm. However, owing to the mechanical imperfections in manufacturing and assembly and the unknown exact point of detection of the electrons within the DLD4444 detector it is essential to calibrate the exact value of $d$ from the energy distribution measurements.

$V_{\text{off}}$ is essentially the work function of the sample relative to that of the detector. A rough value of this can be obtained increasing the voltage on the sample (while adjusting the voltage on the shield according to equation 14) until the electrons stop reaching the detector. This voltage provides a rough estimate of the work function difference between the sample and the detector as the front face of the detector is grounded. However the uncertainty in $V_{\text{off}}$ obtained from this technique is a couple of 100 meV due to the energy spread of the electrons and due to the fact that the electron spot on the detector becomes larger than the detector size at very small values of the effective accelerating voltage $V_e$, making detection of all emitted electrons impossible. A value of $V_{\text{off}} \approx 400$ meV was obtained using this technique.

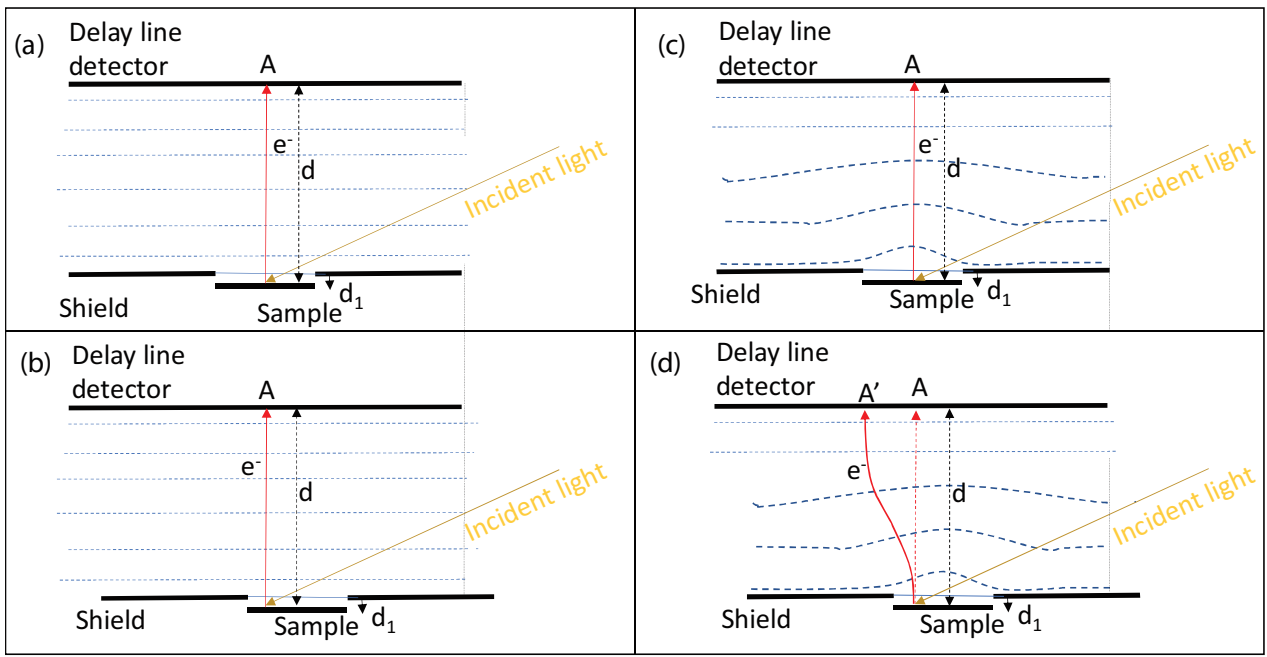

FIG. 6. Illustration of technique used to calculate the voltage offset between the sample and the shield to obtain uniform electric fields. (a) Equipotential lines are parallel to the detector when the voltage offset is equal to the work function difference (b) When equipotential lines are parallel to the detector the electron spot on the detector remains at the same location (A) when the sample-shield assembly is moved w.r.t the detector. Note that the location of incident light is unchanged w.r.t the detector. (c) Equipotential lines are curved close to the sample-shield assembly when the voltage offset is not equal to the work function difference. (d) When equipotential lines are curved the electron spot on the detector changes position (from A to A') when the sample-shield assembly is moved w.r.t the detector.

Below, we demonstrate another technique that allows us to determine both $d$ and $V_{\text{off}}$ to a much better accuracy using a known feature in the longitudinal or total energy distributions. The surface state on the Ag(111) surface is one such feature which manifests itself as a peak in the longitudinal (and total) energy distribution[20].

In our case, the longitudinal energy distributions obtained from the time-of-flight measurements using equation 3 need to satisfy two conditions of physicality:

- the location of the peak corresponding to the surface state in the longitudinal energy distribution (and more generally the entire longitudinal energy distribution) must remain invariant with the voltage $V$ applied to the sample during the measurement (assuming an appropriate corresponding voltage given by equation 14 is applied to the shield to ensure uniform electric field).

- the surface state peak in the longitudinal energy should change by an equivalent amount with a change in photon energy.

When calculating the longitudinal energy distributions from the time-of-flight ($t$) measurements, both the above conditions are satisfied for a unique combination of the values of $d$ and $V_{\text{off}}$.

Figure 7a shows the temporal distributions measured for two voltages $V = -1$ V and $-2$ V at two photon energies of 4.660 eV and 4.605 eV. Longitudinal energy distributions are calculated from these temporal distributions for various values of $d$ around $d = 40$ mm and $V_{\text{off}}$ around $V_{\text{off}} = 400$ mV. The difference of the energy corresponding to the peaks in the longitudinal energy distributions calculated from the temporal distributions measured at the two voltages at the photon energy of 4.660 eV is shown in figure 7b for various values of $d$ and $V_{\text{off}}$. We see that the difference becomes nearly zero along a line in the $d$-$V_{\text{off}}$ plane. All values of $d$ and $V_{\text{off}}$ on this line satisfy the first condition of physicality.

Figure 7c shows the difference of the energies corresponding to the peaks of the longitudinal energy distributions for the two photon energies minus the difference between the two photon energies (55 meV) for various values of $d$ and $V_{\text{off}}$. We see that the difference also becomes nearly zero along a line in the $d$-$V_{\text{off}}$ plane. All values of $d$ and $V_{\text{off}}$ on this line satisfy the second condition of physicality.

Both the conditions are satisfied at the intersection of these two lines which happens at $d = 40.5$ mm and $V_{\text{off}} = 405$ mV. Figure 7d shows the longitudinal energy distributions calculated from the temporal distributions in figure 7a using the values of $d = 40.5$ mm and $V_{\text{off}} = 405$ mV.

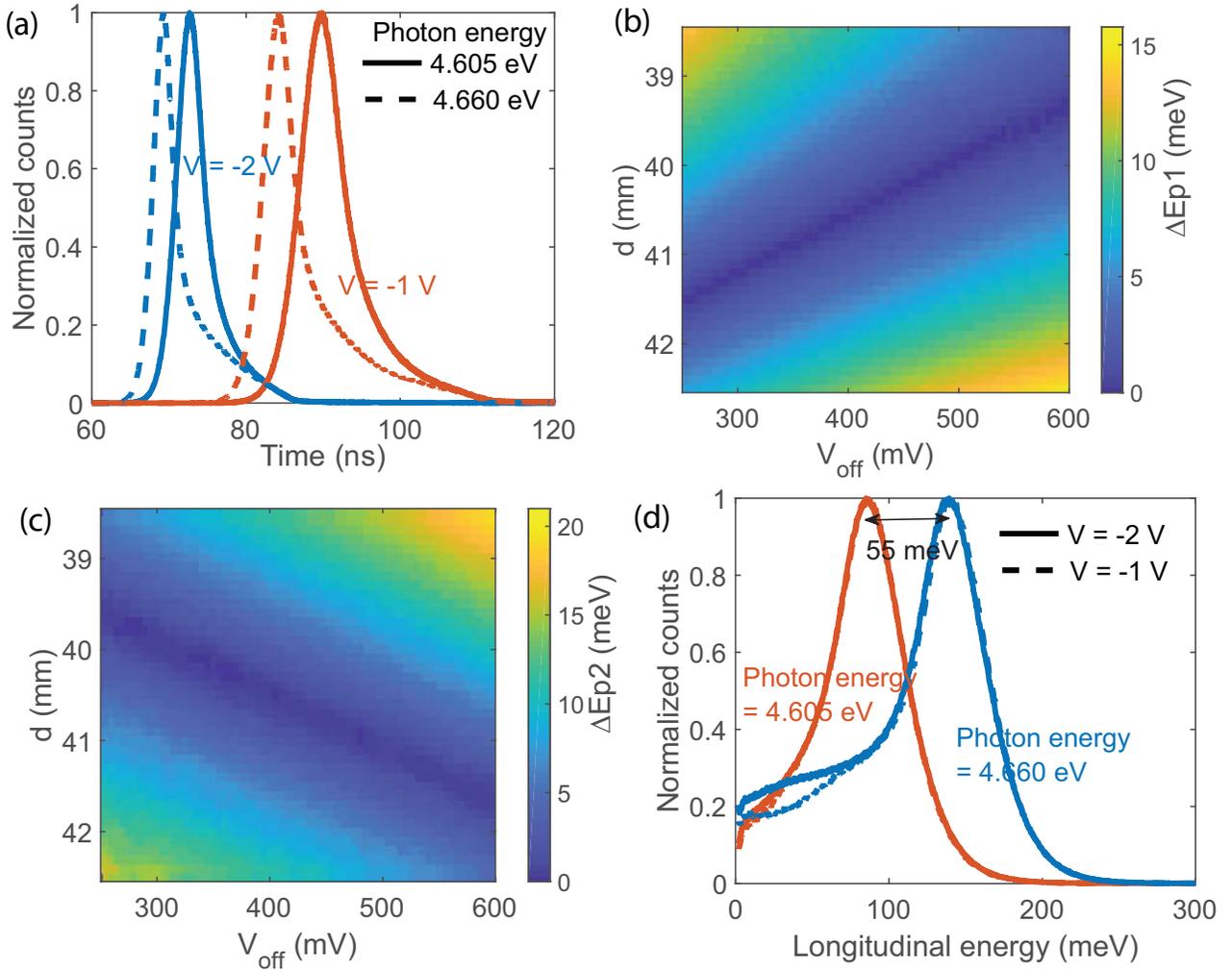

FIG. 7. (a) Temporal distributions measured for two voltages $V = -1$ V and $-2$ V at two photon energies of 4.660 eV and 4.605 eV. (b) The difference of the energy corresponding to the peaks in the longitudinal energy distributions calculated from the temporal distributions measured at the two voltages at the photon energy of 4.660 eV for various values of $d$ and $V_{\text{off}}$ (c) The difference of the energies corresponding to the peaks of the longitudinal energy distributions for the two photon energies minus the difference between the two photon energies (55 meV) for various values of $d$ and $V_{\text{off}}$ (d) Longitudinal energy distributions calculated from the temporal distributions in figure 7a using the values of $d = 40.5$ mm and $V_{\text{off}} = 405$ mV. Both the conditions of physicality are satisfied at these values of $d$ and $V_{\text{off}}$. Note that the longitudinal energy distributions are the same for both the voltages as expected. For the photon energy of 4.660 eV there is a small discrepancy between the longitudinal energy distributions obtained at the two voltages for small longitudinal energies. This is because some of the low energy and high transverse momentum electrons went off screen during the measurement and were not recorded.

Here we used a feature (the surface state on the Ag(111) surface) in the longitudinal energy distribution to obtain the values of $d$ and $V_{\text{off}}$ accurately. In the absence of such a feature, one may similarly use the Fermi energy in the total energy distribution for obtaining the exact values of $d$ and $V_{\text{off}}$.

### C. 3-D energy momentum distributions

Upon obtaining the accurate values of $d$ and $V_{\text{off}}$, we can calculate the 3-D energy and transverse momentum distribution (or equivalently the energy resolved transverse momentum distribution) using equations 1-4 from the measured 3D $x$, $y$ and $t$ distributions. Various cross-sections of the $k_x$-$k_y$-$E$ distributions are shown in figure 8.

Figure 8a shows the $k_y = 0$ cross-section for the $E$ vs $k_x$ distributions for photon energy of 4.660 eV. This cross-section shows the parabolic 2-D Shockley surface state. A parabola can be fitted to the maximum intensity

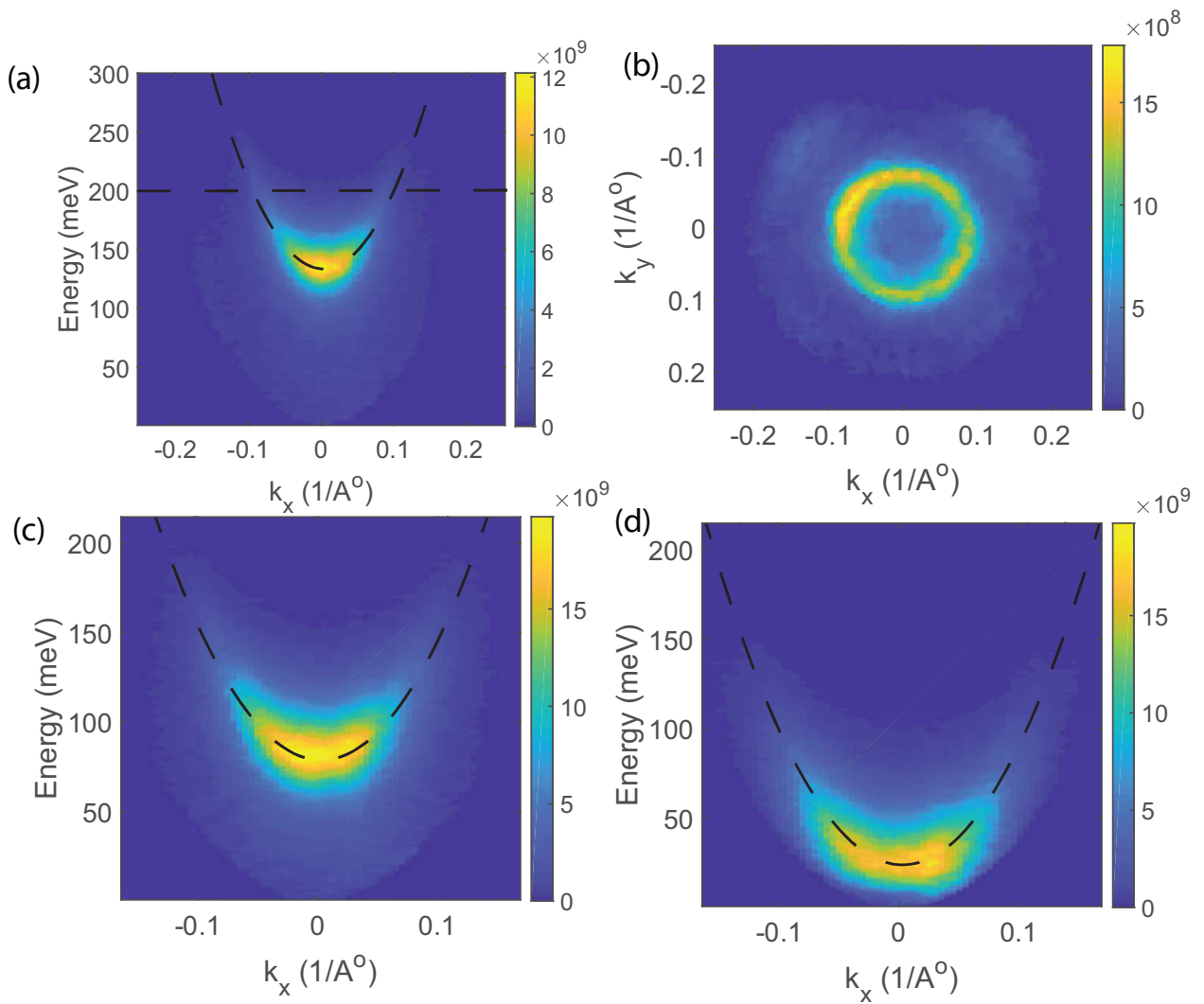

FIG. 8. (a) The $k_y = 0$ cross-section for the energy-transverse momentum distributions for photon energy of 4.660 eV. (b) The $k_x$-$k_y$ cross-section for an energy of $E = 200$ meV for a photon energy of 4.660 eV. The $k_y = 0$ cross-section for the energy-transverse momentum distributions for photon energies of (c) 4.605 eV and (d) 4.550 eV. The energy of the emitted surface state electrons goes down with photon energy as expected.

points for each value of $k_x$. The effective mass computed from this parabola is found to be $0.53m_e$ and is in good agreement with previous results at room temperature[21]. Figure 8b shows the $k_x$-$k_y$ cross-section for an energy of $E = 200$ meV for a photon energy of 4.660 eV. This shows a ring in the momentum distribution due to the 2-D surface state. Figures 8c and 8d show the $k_y = 0$ cross-section for the E vs $k_x$ distributions for photon energies of 4.605 eV and 4.550 eV along with the same parabolic fit used for the 4.660 eV case. Within the resolution of the system (this includes the band broadening due to thermal effects at room temperature and photon energy spread of the laser along with the resolution of the instrument) the surface state is identical except for a shift in energy that corresponds to the changing photon energy, demonstrating the ability of this instrument to measure meV energy scale photoelectrons accurately. All measurements shown in figure 8 we performed at $V = -1$

V.

### D. Measurement of MTE

The MTE is a crucial figure of merit for photoemission based electron sources. As shown in section 2, the time-of-flight setup can give very accurate transverse energy distributions that do not require the knowledge of $d$ and $V_{\text{off}}$ and are relatively robust towards stray fields. This makes the time-of-flight an excellent technique to measure the MTE. Figure 9 shows the MTE obtained as a function of photon energy. These measurements agree well with previous measurements performed using other techniques[11]. We see that the MTE first reduces with photon energy, reaches a minimum and then increases again. At threshold photon energies, we are exciting electrons from the Fermi level, where the surface state has a significant transverse momentum. As the photon energy is increased, electrons closer to the bottom of the surface state band with lower transverse momentum are emitted. At the binding energy of the surface state, electrons are emitted with zero transverse momentum. Thus as we increase the photon energy from threshold by a value of the surface state energy, MTE reduces to a minimum value and then increases again as high transverse momentum bulk states at the Fermi level start to emit.

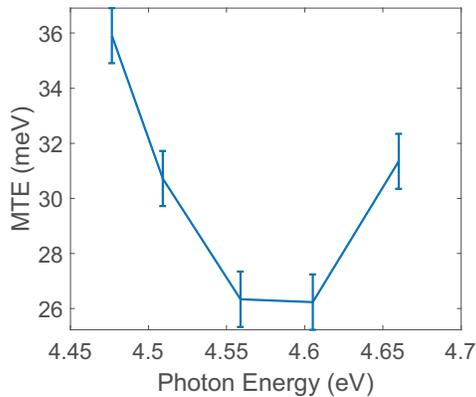

FIG. 9. MTE obtained as a function of photon energy for the Ag(111) surface.

## IV. CONCLUSION

In this paper we have reported the design and development of a 3-D energy-momentum analyzer based on the time-of-flight measurement. This analyzer can obtain sub-meV resolution while being capable of measuring meV-scale energy electrons. Such an analyzer will be an ideal tool for investigating photoemission processes close to the photoemission threshold and aid in the development of electron sources with very small energy and transverse momentum spreads. We have presented a detailed analysis of all possible sources of error and concluded that one should be able to achieve a sub-meV resolution in both the longitudinal and transverse energy. Using the example of the Ag(111) surface state, we have also demonstrated the operation of this analyzer and discussed variaous techniques to calibrate to unknown parameters required for accurate measurements of the longitudinal and transverse energies and momenta.

In practice demonstrating a sub-meV energy resolution requires a very sharp feature in the energy distribution and a very narrow band-width laser. Sharp features in the energy distribution can be obtained at cryogenic (LHe) temperatures and the narrow band-width can be obtained by using picosecond laser pulses. In this case, this was done by dispersing light from the 150 fs Ti Sapphire laser oscillator and energy selecting with a slit. Such a reduction in bandwidth results in pulse length broadening. With the implementation of these two, it should be possible to demonstrate the sub-meV resolution of this analyzer at sub-100 meV electron energies.

This work was supported by the U.S. National Science Foundation under Award No. PHY-1549132, the Center for Bright Beams and by the Director, Office of Science, Office of Basic Energy Sciences of the U.S. Department of Energy, under Contracts No. KC0407-ALSJNT-I0013 and No. DE-AC02-05CH11231. We would also like to acknowledge the useful discussions with Dr. Andreas Olsner regarding the operation of the delay line detector.

[1] S. Hufner, *Photoelectron Spectroscopy Principles and Applications* (Springer (Berlin), 2003).
[2] W. Zhang, *Photoemission Spectroscopy on High Temperature Superconductor* (Springer-Verlag Berlin Heidelberg, 2013).
[3] A. Vollmer, R. Ovsyannikova, M. Gorgoi, S. Krausea, M. Oehzelt, A. Lindbladb, N. Martenssonb, S. Svenssonb, P. Karlssonc, M. Lundvuist, T. Schmeiler, J. Pflaumd, and N. Koch, J. Elec. Spectr. Rel. Phenom. **185**, 55 (2012).
[4] https://www.scientaomicron.com/en/products/electron-spectrometer.
[5] J. D. Koralek, J. F. Douglas, N. C. Plumb, and J. D. Griffith, Rev. Sci. Instrum. **78**, 053905 (2007).
[6] A. H. Zewail, Annu. Rev. Phys. Chem. **57**, 65 (2006).
[7] S. M. Gruner, D. Bilderback, I. Bazarov, K. Finkelstein, G. Krafft, L. Merminga, H. Padamsee, Q. Shen, C. Sinclair, and M. Tigner, Rev. Sci. Instrum. **73**, 1402 (2002).
[8] P. Emma, R. Akre, J. Arthur, R. Bionta, C. Bostedt, J. Bozek, A. Brachmann, P. Bucksbaum, R. Coffee, F. Decker, *et al.*, Nat. Photon. **4**, 641 (2010).
[9] P.Musumeci, J. Navarro, J.B.Rosenzweig, L.Cultrera, I.Bazarov, J.Maxson, S.Karkare, and H.Padmore, Nucl. Instrum. Meth. A. **907**, 209 (2018).
[10] S. Karkare, L. Boulet, L. Cultrera, B. Dunham, X. Liu, W. Schaff, and I. Bazarov, Phys. Rev. Lett. **112**, 097601 (2014).
[11] S. Karkare, J. Feng, X. Chen, W. Wan, F. J. Palomares, T.-C. Chiang, and H. A. Padmore, Phys. Rev. Lett. **118**, 164802 (2017).
[12] J. Feng, J. Nasiatka, W. Wan, T. Vecchione, and H. A. Padmore, Rev. Sci. Instrum. **86**, 015103 (2015).
[13] L. B. Jones, H. E. Scheibler, D. V. Gorshkov, A. S. Terekhov, B. L. Militsyn, and T. C. Q. Noakes, J. Appl. Phys. **121**, 225703 (2017).
[14] S. Karkare, L. Cultrera, Y. Hwang, R. Merluzzi, and I. Bazarov, Rev. Sci. Instrum. **86**, 033301 (2015).
[15] D. A. Orlov, M. Hoppe, U. Weigel, D. Schwalm, A. S. Terekhov, and A. Wolf, Appl. Phys. Lett. **78**, 2721 (2001).
[16] L. J. Devlin, L. B. Jones, T. C. Q. Noakes, C. P. Welsch, and B. L. Militsyn, Rev. Sci. Instrum. **89**, 083305 (2018).
[17] P. Kirchmann, L. Rettig, D. Nandi, U. Lipowski, M. Wolf, and U. Bovensiepen, Appl. Phys. A **91**, 211 (2008).
[18] http://www.surface-concept.com/.
[19] https://laacg.lanl.gov/laacg/services/download_sf.phtml.
[20] S. D. Kevan and W. Eberhardt, *Angle resolved photoemission: theory and current applications* (Elsevier, Amsterdam, 1992).
[21] S. Kevan and R. Gaylord, Phys. Rev. B **36**, 5809 (1987).